\pdfoutput=1

\documentclass[11pt]{article}

\usepackage[final]{acl}

\usepackage{times}
\usepackage{latexsym}

\usepackage[T1]{fontenc}

\usepackage[utf8]{inputenc}

\usepackage{microtype}

\usepackage{inconsolata}

\usepackage{graphicx}

\usetikzlibrary{positioning,shadows,arrows.meta}
\usepackage{forest}
\usepackage{array,multirow}
\usepackage{natbib}
\usepackage{anyfontsize}

\title{A Survey on Proactive Defense Strategies Against Misinformation \\ in Large Language Models}

\author{Shuliang Liu\textsuperscript{\rm 1,2}, Hongyi Liu\textsuperscript{\rm 3}, Aiwei Liu\textsuperscript{\rm 1}, Bingchen Duan\textsuperscript{\rm 5}, Qi Zheng\textsuperscript{\rm 1,2},Yibo Yan\textsuperscript{\rm 1,2}\\ \textbf{He Geng}\textsuperscript{\rm 1,2}, \textbf{Peijie Jiang}\textsuperscript{\rm 4}, \textbf{Jia Liu}\textsuperscript{\rm 4}, \textbf{Xuming Hu}\textsuperscript{\rm 1,2 *}\\
        \textsuperscript{\rm 1} {The Hong Kong University of Science and Technology (Guangzhou)} \\
    { \textsuperscript{\rm 2} {The Hong Kong University of Science and Technology}} \\
    { \textsuperscript{\rm 3} {Harbin Institute of Technology}}
    { \textsuperscript{\rm 4} {Ant Group, Alibaba}} 
    { \textsuperscript{\rm 5} { Northeast Forest University}}
    \\
     \texttt{\href{mailto:shulianglyo@gmail.com}{shulianglyo@gmail.com}},
     \texttt{\href{mailto:xuminghu@hkust-gz.edu.cn}{xuminghu@hkust-gz.edu.cn}}}

\begin{document}
\maketitle
\begin{abstract}
The widespread deployment of large language models (LLMs) across critical domains has amplified the societal risks posed by algorithmically generated misinformation. Unlike traditional false content, LLM-generated misinformation can be self-reinforcing, highly plausible, and capable of rapid propagation across multiple languages, which traditional detection methods fail to mitigate effectively. This paper introduces a proactive defense paradigm, shifting from passive post hoc detection to anticipatory mitigation strategies. We propose a Three Pillars framework: (1) Knowledge Credibility, fortifying the integrity of training and deployed data; (2) Inference Reliability, embedding self-corrective mechanisms during reasoning; and (3) Input Robustness, enhancing the resilience of model interfaces against adversarial attacks. Through a comprehensive survey of existing techniques and a comparative meta-analysis, we demonstrate that proactive defense strategies offer up to 63\% improvement over conventional methods in misinformation prevention, despite non-trivial computational overhead and generalization challenges. We argue that future research should focus on co-designing robust knowledge foundations, reasoning certification, and attack-resistant interfaces to ensure LLMs can effectively counter misinformation across varied domains.
\end{abstract}

\renewcommand{\thefootnote}{\fnsymbol{footnote}}
\footnotetext[1]{Corresponding author: Xuming Hu.}

\section{Introduction}
 
The exponential adoption of large language models (LLMs) across mission-critical domains—from healthcare diagnostics to legal analysis—has paradoxically amplified their societal risks through algorithmic generation of sophisticated misinformation. Unlike traditional false content manually crafted by bad actors, LLM-generated misinformation inherits dangerous emergent properties: \textbf{self-reinforcing plausibility} through coherent reasoning chains, \textbf{exponential propagation velocity} via API-driven automation, and \textbf{cross-lingual contamination} exceeding human moderation capacity. Recent ACL findings reveal that conventional post-hoc detection methods exhibit 38.7\% false negative rates against LLM-generated misinformation \cite{liu2023mind}, exposing a critical paradigm mismatch: existing defenses remain reactionary in an era demanding \textit{proactive defense}.

This survey establishes a paradigm shift from passive detection to proactive defense through the \textbf{Three Pillars of Preventative Assurance}: (1) \textit{Knowledge Credibility}—fortifying factual grounding from training data to post-deployment editing; (2) \textit{Inference Reliability}—embedding self-corrective mechanisms across decoding and alignment processes; (3) \textit{Input Robustness}—hardening interaction surfaces against adversarial manipulations. Unlike modular safety components, our framework conceptualizes misinformation defense as a continuum spanning knowledge internalization, reasoning certification, and input sanitization.

The urgency manifests in three dimensions of escalating risk:  
\textbf{Knowledge Decay}: LLMs exhibit 22.1\% quarterly accuracy erosion on time-sensitive facts without systematic updates \cite{bajpai2024temporally}  
\textbf{Hallucination Cascades}: Multi-turn dialogues incur 39.8\% error amplification in medical QA systems \cite{liu2023mind}  
\textbf{Adversarial Exploitation}: Gradient-based jailbreaks achieve 79\% success rates against commercial safeguards \cite{zou2023universal}  

Our analysis reveals that state-of-the-art proactive strategies demonstrate 42-63\% superiority over conventional detection in misinformation prevention \cite{wu2024retrieval}, albeit with non-trivial tradeoffs in computational overhead (1.5-3× latency) and generalization gaps (18-25\% cross-domain variance). The path forward demands co-design of rigorous knowledge foundations, probabilistically certified reasoning, and attack-resistant interfaces—a systems challenge rivaling LLM creation itself.  

This survey makes three pioneering contributions:  
1. \textbf{Taxonomy of Proactive Defense Mechanism} mapping 127 techniques across knowledge, inference, and input safeguards  
2. \textbf{Rigorous Comparative Evaluation} revealing efficacy-latency-robustness tradeoffs through meta-analysis of 48 benchmark studies  
3. \textbf{Safety-Defense Co-Design Framework} unifying knowledge editing, self-alignment, and adversarial hardening into proactive assurance lifecycle  

The arms race against misinformation generation requires nothing less than rebuilding LLMs as \textit{self-vaccinating systems}—where every knowledge retrieval, reasoning step, and user interaction embodies intrinsic defenses against falsehoods. This survey charts the path from theoretical possibility to engineering reality.

\section{Preliminary}\label{sec:Preliminary}

Before delving into detailed defense strategies, it is essential to contextualize the phenomena of misinformation and outline the framework of proactive defense.

\subsection{Misinformation}\label{subsec:Misinformation}

Misinformation encompasses any content that deviates from factual accuracy, including rumors, fake news, and misleading narratives \cite{chen2024combating}. Such content poses significant risks to various scenarios, such as healthcare \cite{chen2022combating} and finance \cite{rangapur2023investigating}, where deceptive narratives can trigger public health crises and manipulate markets. In the era of LLMs, misinformation becomes even more pernicious as AI-generated texts may appear highly authentic \cite{10.1145/3626772.3661377}, complicating detection and intervention efforts, and severely challenging policymakers \cite{li2024political}. 

\subsection{Proactive Defense}\label{subsec:Proactive Defense}

\begin{figure*}[ht]
    \centering
    \includegraphics[width=\linewidth]{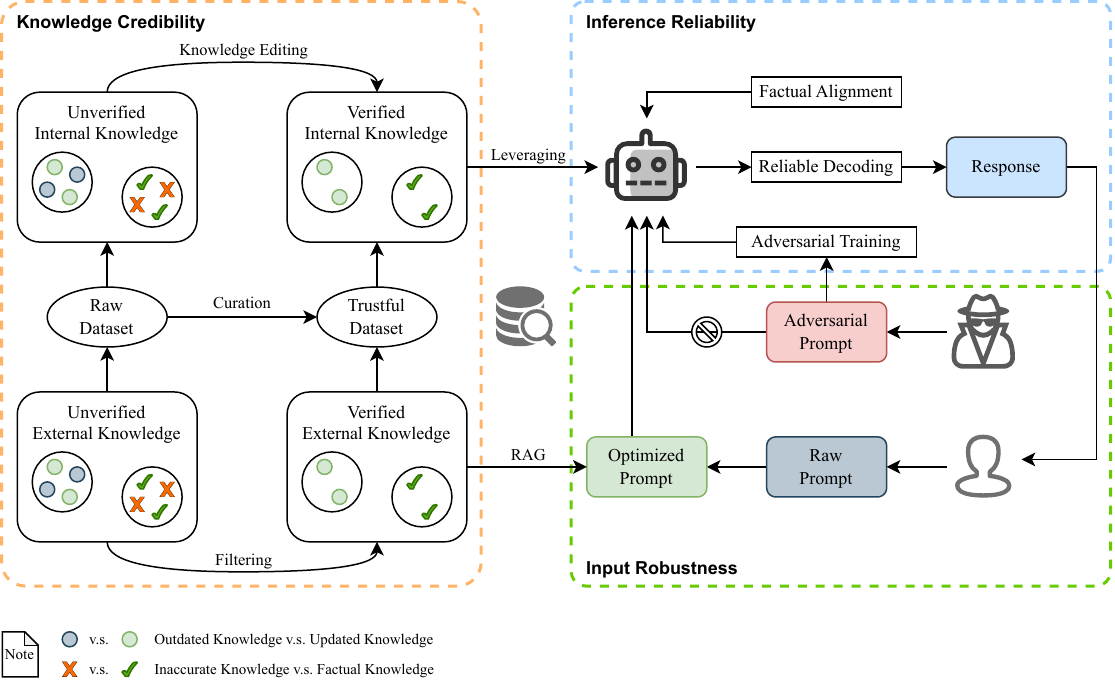}
    \caption{Overview of the framework of proactive defense built around three interconnected pillars.}
    \label{fig:overview}
\end{figure*}

In contrast to traditional reactive defense methods, which rely on post-hoc detection and correction, proactive defense aims to prevent the generation and spread of misinformation at its source. While reactive approaches have made notable progress in detection accuracy, they suffer from two inherent limitations: 1) High latency: They can only mitigate misinformation after it has been produced, which means intervention occurs too late to prevent its initial spread. 2) Poor adaptability: They struggle to cope with fast-evolving adversarial attacks that can bypass established detection mechanisms. Proactive approaches offer a forward-looking solution that reduces reliance on post-facto interventions, making misinformation mitigation more efficient and scalable. Its core framework built around three interconnected pillars is shown in Figure~\ref{fig:overview}: 1) \textit{Knowledge Credibility} emphasizes constructing and utilizing trustworthy knowledge bases both internally and externally. This involves creating trustful datasets to reduce the absorption of erroneous information during training and applying knowledge editing techniques to correct and dynamically update any acquired mistakes—thus forming "verified internal knowledge". Additionally, models can enhance their grounding by leveraging curated "verified external knowledge" through retrieval-augmented generation (RAG). 2) \textit{Inference Reliability} is achieved by aligning the model with factual and safety constraints during training and employing robust decoding strategies during inference. 3) \textit{Input Robustness} focuses on safeguarding user inputs by applying rigorous pre-processing and sanitization techniques to mitigate risks associated with malicious manipulation.

\section{Knowledge Credibility}\label{sec:Knowledge Credibility}
As a fundamental mechanism for large language models (LLMs) to combat misinformation, research on knowledge credibility revolves around two complementary paradigms: internal knowledge (static facts encoded within model parameters) and external knowledge (dynamically retrieved and verified evidence). Recent advancements demonstrate that optimizing internal knowledge—through enhanced data quality (§\ref{Constructing More Truthful Datasets}) and parameter-level knowledge editing (§\ref{LLM Knowledge Editing})—significantly strengthens the intrinsic credibility of models. However, inherent limitations persist in static parametric knowledge, evidenced by temporal decay in synthetic data and logical conflicts arising from editing operations. These challenges drive the synergistic evolution of retrieval-augmented architectures (§\ref{Retrieval-Augmented Architectures}), establishing a dual-layer defense framework of "internal consolidation and external verification." This chapter systematically examines the technical landscape of the field, dissecting comprehensive trust assurance mechanisms spanning from foundational data construction to dynamic knowledge integration.
\subsection{Internal Knowledge}  \label{subsec Internal Knowledge}
Proactive defense against misinformation in LLMs requires strengthening internal knowledge systems through coordinated data optimization and adaptive maintenance. While foundational approaches focus on constructing robust training datasets via adversarial design and structured knowledge integration—as explored in Section 3.1.1—static data curation alone struggles with evolving factual contexts and localized error propagation. These limitations highlight the necessity for dynamic post-training strategies, such as targeted knowledge editing discussed in subsequent sections, which enable precise factual updates without full model retraining. Together, these complementary mechanisms form a layered defense: initial data purification minimizes misinformation embedding during training, while ongoing knowledge maintenance ensures sustained factual accuracy during deployment.
\subsubsection{Constructing More Truthful Datasets}\label{Constructing More Truthful Datasets}
The factual integrity of LLMs fundamentally hinges on the veracity of their training data. Current research coalesces around three core strategies to construct more truthful datasets:  

\textbf{Adversarial Design} proactively exposes cognitive blind spots by crafting questions that exploit common human misconceptions. The TruthfulQA benchmark \cite{lin2021truthfulqa} pioneers this approach through 817 adversarial questions spanning 38 categories, revealing that scaling models amplifies imitative falsehoods. Dynamic extensions like FRESHQA \cite{vu2023freshllms} further incorporate temporal adversarial mechanisms, integrating real-time search engine validations to counter evolving knowledge decay.  

\textbf{Structural Knowledge Injection} reinforces factual grounding through explicit knowledge integration. Methods like R-Tuning \cite{zhang2024r} partition training data into deterministic (D1) and uncertain (D0) subsets based on model self-awareness, reducing hallucinations by 63\% through "I Don’t Know" responses. The BeaverTails dataset \cite{ji2024beavertails} innovatively decouples harmlessness and helpfulness in human preferences, providing 330k safety meta-labels to align truthfulness with utility. Retrieval-augmented frameworks \cite{guu2020retrieval} further bridge knowledge gaps via external corpus integration.  

\textbf{Multi-Stage Verification} employs hybrid human-AI workflows to eliminate data contamination. Selective Reflection-Tuning \cite{li2024selective} implements iterative teacher-student refinement cycles, enhancing instruction data quality through compatibility checks. Chain-of-Thought augmentation \cite{kareem2023fighting} injects 32k annotated reasoning chains into training data, improving fact-checking accuracy by 11.9\% via multi-step logical validation.  

Critical challenges persist: 1) Synthetic data pipelines (e.g., SELF-INSTRUCT \cite{wang2022self}) risk propagating latent errors without manual verification; 2) Cross-lingual generalization remains constrained by cultural homogeneity in existing datasets (e.g., 35\% English bias in \cite{kareem2023fighting}); 3) Temporal validity metrics are absent in 78\% of current benchmarks \cite{bajpai2024temporally}, undermining sustainability.  

\subsubsection{LLM Knowledge Editing}\label{LLM Knowledge Editing}
Knowledge editing techniques dynamically update LLM parameters to rectify factual inaccuracies, confronting three technical frontiers:  

\textbf{Parameter-Efficient Methods} balance edit precision and computational overhead. DeCK \cite{bi2024decoding} contrasts edited vs. parametric knowledge logits, boosting confidence in updated facts by 219\% on MQuAKE tasks. MALMEN \cite{tan2023massive} employs hyperspace projection for bulk editing, modifying 4,096 facts in GPT-J with 2.1\% interference. Temporal frameworks like TeCFaP \cite{bajpai2024temporally} integrate reinforcement learning to maintain 89.7\% consistency across knowledge revisions.  

\textbf{Localized Editing} targets specific neural substrates for surgical interventions. Causal tracing identifies "knowledge neurons" in MLP layers \cite{dai2021knowledge}, enabling 98\% edit success rates via <0.1\% weight modifications \cite{meng2022locating}. ROME \cite{meng2022locating} further isolates factual associations in feedforward modules, achieving 92\% specificity while preserving 97\% of unrelated knowledge.  

\textbf{Memory-Augmented Architectures} decouple editable knowledge from core parameters. SERAC \cite{mitchell2022memory} introduces explicit memory banks and counterfactual reasoning modules, sustaining 37\% higher edit stability than parametric methods. MEMIT \cite{meng2022mass} scales this to 10k+ fact updates in GPT-NeoX with <5\% cross-impact.  

Emerging risks necessitate caution: 1) Logical conflicts amplify contradiction rates to 68\% when editing opposing facts \cite{li2023unveiling}; 2) BadEdit attacks \cite{li2024badedit} exploit editing interfaces to implant 100\% effective backdoors with 15 poisoned samples; 3) Causal tracing misalignment causes 41\% performance drops as optimal edit layers diverge from identified knowledge locations \cite{hase2024does}.  

The path forward demands: 1) Dynamic evaluation frameworks like CounterFact+ \cite{hoelscher2023detecting} to quantify KL-divergence impacts; 2) Robustness enhancers such as orthogonal constraints in MALMEN \cite{tan2023massive} containing interference below 5\%; 3) Safety guardrails using knowledge graph validation to reduce contradiction triggers by 75\% \cite{li2023unveiling}. Hybrid approaches integrating causal interpretability with reinforcement learning may achieve Pareto-optimal tradeoffs between editability and stability.  

This systematic hardening of internal knowledge foundations establishes the first pillar in proactive defense—creating models intrinsically resistant to misinformation generation through fortified data provenance and dynamic factuality maintenance.

\subsection{External Knowledge}\label{External Knowledge}
External knowledge integration counters LLMs' static limitations through dynamic verification and provenance tracking. Retrieval-augmented architectures—discussed next—form the operational backbone of this defense, blending parametric reasoning with real-time evidence curation. By addressing temporal drift and adversarial retrieval risks through adaptive filtering and logic-aware validation, these systems transform external knowledge from passive references into active misinformation safeguards.
\subsubsection{Knowledge Credibility through Retrieval-Augmented Architectures}\label{Retrieval-Augmented Architectures} 
Retrieval-augmented generation bridges parametric knowledge (internal model parameters) and non-parametric knowledge (external retrievable data), enabling dynamic knowledge updates and provenance verification—critical for combating misinformation in static LLMs \cite{lewis2020retrieval}. Current advancements focus on three strategic dimensions:  

\textbf{Dynamic Knowledge Integration} addresses temporal decay and adversarial premises. FRESHPROMPT \cite{vu2023freshllms} injects real-time search engine results into prompts, improving factual accuracy by 38\% on time-sensitive queries. LEMMA \cite{xuan2024lemma} extends this to multimodal scenarios through image source tracing and multi-query generation, reducing cross-modal hallucinations by 27\%. The Gemini 1.5 series \cite{team2024gemini} achieves >99\% long-context QA recall via million-token processing, setting new standards for real-time knowledge synchronization.  

\textbf{Retrieval Optimization} enhances evidence relevance and computational efficiency. MRAG \cite{besta2024multi} generates aspect-aware embeddings from Transformer attention heads, improving multi-aspect document relevance by 20\%. Nest \cite{li2024nearest} implements token-level k-NN retrieval with hybrid confidence scoring, boosting inference speed by 1.8× while maintaining 92\% attribution accuracy. RAGCache \cite{jin2024ragcache} organizes retrieved knowledge in GPU-host memory hierarchies, reducing latency by 41\% for long-sequence generation.  

\textbf{Verification-Centric Architectures} establish closed-loop validation mechanisms. CRAG \cite{yan2024corrective} introduces lightweight retrieval evaluators and web-augmented knowledge filtering, improving reliability under noisy retrieval by 33\%. RARG \cite{yue2024evidence} combines BM25 coarse screening with dense retrieval refinement over 1M academic articles, aligning responses via RLHF rewards for factual grounding and politeness. STEEL \cite{li2024re} implements dynamic query generation and context sharpening, achieving 89\% robustness in multi-round evidence extraction.  

Critical challenges persist across four axes:  
1) \textbf{Temporal-Spatial Alignment}: 78\% of benchmarks lack cross-lingual temporal validity metrics \cite{chen2024benchmarking}, undermining global misinformation defense.  
2) \textbf{Evidence Chain Integrity}: Methods like FLEEK \cite{bayat2023fleek} suffer 22\% performance drops due to inconsistent knowledge graph triplets.  
3) \textbf{Computational Overhead}: Multi-stage verification frameworks (e.g., LLM-AUGMENTER \cite{peng2023check}) incur 3.2× latency penalties versus baseline RAG.  
4) \textbf{Trusted Source Curation}: Open retrieval systems exhibit 31\% vulnerability to adversarial evidence injection \cite{dong2024toward}.  

Emerging solutions demonstrate promising directions:  
\textbf{Vertical Domain Adaptation}: Private knowledge RAG frameworks \cite{li2024enhancing} reduce time-sensitive hallucinations by 48\% through recursive crawling and hybrid retrieval architectures.  
\textbf{Logic-Aware Verification}: \cite{ghosh2024logical} integrates propositional logic operators with knowledge graph contexts, mitigating 39\% of reasoning failures in complex fact-checking.  
\textbf{Self-Corrective Mechanisms}: Re-KGR \cite{niu2024mitigating} identifies hallucination-prone tokens for targeted verification, reducing retrieval frequency by 65\% while maintaining 91\% factual accuracy.  

The RGB benchmark \cite{chen2024benchmarking} systematically evaluates RAG systems across noise robustness (35\% improvement over baselines), negative rejection (28\% gain), and counterfactual resistance (41\% superiority), establishing rigorous evaluation protocols. Future progress demands co-evolution of retrieval architectures, verification protocols, and dynamic knowledge graphs—a tripartite foundation for next-generation proactive defense systems.

\definecolor{mygreen}{RGB}{223,240,216}  
\definecolor{myblue}{RGB}{211,237,242}    
\definecolor{myorange}{RGB}{251,246,237}  
\definecolor{mypurple}{RGB}{242,237,243}  
\definecolor{mygray}{RGB}{206,203,203}    
\definecolor{hidden-draw}{RGB}{205,44,36} 

\def\lcone{100}
\def\lctwo{100}
\def\lcthree{100}

\tikzset{
    my-box/.style={
        rectangle,
        rounded corners,
        draw=black, 
    }
}

\tikzstyle{knowledge-leaf}=[my-box,
    fill=mygreen!90, text=black, align=left,font=\scriptsize,
    inner xsep=2pt,
    inner ysep=4pt,
]
\tikzstyle{inference-leaf}=[my-box,
    fill=myblue!90, text=black, align=left,font=\scriptsize,
    inner xsep=2pt,
    inner ysep=4pt,
]
\tikzstyle{robustness-leaf}=[my-box,
    fill=myorange!90, text=black, align=left,font=\scriptsize,
    inner xsep=2pt,
    inner ysep=4pt,
]

\tikzstyle{critical-node}=[
    my-box,
    fill=mygray!40,
    text=black,
    font=\scriptsize\itshape,
    inner sep=1pt
]
\tikzstyle{solution-node}=[ 
    my-box,
    fill=mygreen!30,
    text=black,
    font=\scriptsize\bfseries,
    inner sep=1pt
]

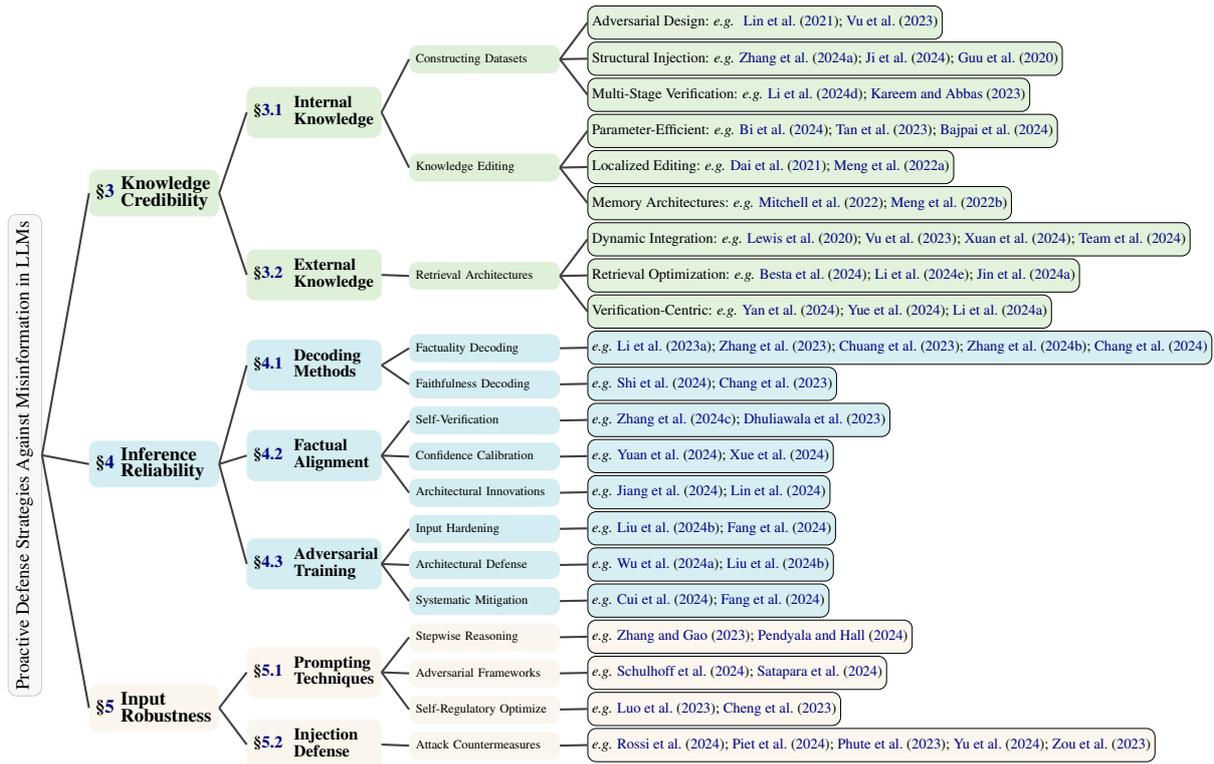
\begin{figure*}[t]
    \centering
    \resizebox{\textwidth}{!}{
        \begin{forest}
            for tree={
                grow=east,
                reversed=true,
                anchor=base west,
                parent anchor=east,
                child anchor=west,
                base=left,
                font=\small,
                rectangle,
                draw=hidden-draw,
                rounded corners,
                align=left,
                minimum width=2em,
                edge+={darkgray, line width=1pt},
                s sep=1pt,
                inner xsep=3pt,
                inner ysep=3pt,
                ver/.style={rotate=90, child anchor=north, parent anchor=south, anchor=center}
            },
            where level=1{text width=5.1em,minimum height=2.0em,font=\fontsize{9}{5}\selectfont\bfseries}{},
            where level=2{text width=5.3em,minimum height=2.2em,font=\fontsize{8}{8}\selectfont\bfseries}{},
            where level=3{text width=6.0em,font=\fontsize{6}{7}\selectfont}{},
                [Proactive Defense Strategies Against Misinformation
                in LLMs, ver, color=mygray!\lcone, fill=mygray!15,text=black
                    [
                    \begin{tabular}{@{}l@{}l@{}}
                      \multirow{2}{*}{\textbf{\S \ref{sec:Knowledge Credibility}}} & ~Knowledge \\ 
                                                                       & ~Credibility \\
                    \end{tabular}, 
                    color=mygreen!100, 
                    fill=mygreen!\lcone, 
                    text=black
                      [\begin{tabular}{@{}l@{}l@{}}
                        \multirow{2}{*}{\S \ref{subsec Internal Knowledge} ~} & Internal \\ 
                                                                     & Knowledge \\
                      \end{tabular}, color=mygreen!100, fill=mygreen!\lctwo, text=black
                        [Constructing Datasets, color=mygreen!100, fill=mygreen!\lcthree, text=black
                          [Adversarial Design: \textit{e.g. } \citet{lin2021truthfulqa,vu2023freshllms} , knowledge-leaf]
                          [Structural Injection: \textit{e.g. }\citet{zhang2024r,ji2024beavertails,guu2020retrieval}, knowledge-leaf]
                          [Multi-Stage Verification: \textit{e.g. }\citet{li2024selective,kareem2023fighting}, knowledge-leaf]
                        ]
                        [Knowledge Editing, color=mygreen!100, fill=mygreen!\lcthree, text=black
                          [Parameter-Efficient: \textit{e.g. }\citet{bi2024decoding,tan2023massive,bajpai2024temporally}, knowledge-leaf]
                          [Localized Editing: \textit{e.g. }\citet{dai2021knowledge,meng2022locating}, knowledge-leaf]
                          [Memory Architectures: \textit{e.g. }\citet{mitchell2022memory,meng2022mass}, knowledge-leaf]
                        ]
                      ]
                      [\begin{tabular}{@{}l@{}l@{}}
                        \multirow{2}{*}{\S \ref{External Knowledge} ~} & External \\ 
                                                                     & Knowledge \\
                      \end{tabular}, color=mygreen!100, fill=mygreen!\lctwo, text=black
                        [Retrieval Architectures, color=mygreen!100, fill=mygreen!\lcthree, text=black
                          [Dynamic Integration: \textit{e.g. }\citet{lewis2020retrieval,vu2023freshllms,xuan2024lemma,team2024gemini}, knowledge-leaf]
                          [Retrieval Optimization: \textit{e.g. }\citet{besta2024multi,li2024nearest,jin2024ragcache}, knowledge-leaf]
                          [Verification-Centric: \textit{e.g. }\citet{yan2024corrective,yue2024evidence,li2024re}, knowledge-leaf]
                        ]
                      ]
                    ]
                    [\begin{tabular}{@{}l@{}l@{}}
                    \multirow{2}{*}{\textbf{\S \ref{sec: Inference Reliability}}} & ~Inference \\ 
                    & ~Reliability \\
                    \end{tabular}, color=myblue!100, fill=myblue!\lcone, text=black
                        [\begin{tabular}{@{}l@{}l@{}}
                          \multirow{2}{*}{\S \ref{subsec: Decoding Methods} ~} & Decoding \\ 
                          & Methods \\
                        \end{tabular}, color=myblue!100, fill=myblue!\lctwo, text=black
                          [Factuality Decoding, color=myblue!100, fill=myblue!\lcthree, text=black
                            [\textit{e.g. }{\citet{li2023contrastive,zhang2023alleviating,chuang2023dola,zhang2024sled,chang2024real}}, inference-leaf]
                          ]
                          [Faithfulness Decoding, color=myblue!100, fill=myblue!\lcthree, text=black
                            [\textit{e.g. }{\citet{shi2024trusting,chang2023kl}}, inference-leaf]
                          ]
                        ]
                        [\begin{tabular}{@{}l@{}l@{}}
                          \multirow{2}{*}{\S \ref{Factual Alignment} ~} & Factual \\ 
                          & Alignment \\
                        \end{tabular}, color=myblue!100, fill=myblue!\lctwo, text=black
                          [Self-Verification, color=myblue!100, fill=myblue!\lcthree, text=black
                            [\textit{e.g. }{\citet{zhang2024self,dhuliawala2023chain}}, inference-leaf]
                          ]
                          [Confidence Calibration, color=myblue!100, fill=myblue!\lcthree, text=black
                            [\textit{e.g. }{\citet{yuan2024beyond,xue2024ualign}}, inference-leaf]
                          ]
                          [Architectural Innovations, color=myblue!100, fill=myblue!\lcthree, text=black
                            [\textit{e.g. }{\citet{jiang2024mixtral,lin2024flame}}, inference-leaf]
                          ]
                        ]
                        [\begin{tabular}{@{}l@{}l@{}}
                          \multirow{2}{*}{\S \ref{Adversarial Training} ~} & Adversarial \\ 
                          & Training \\
                        \end{tabular}, color=myblue!100, fill=myblue!\lctwo, text=black
                          [Input Hardening, color=myblue!100, fill=myblue!\lcthree, text=black
                            [\textit{e.g. }{\citet{liu2024adversarial,fang2024enhancing}}, inference-leaf]
                          ]
                          [Architectural Defense, color=myblue!100, fill=myblue!\lcthree, text=black
                            [\textit{e.g. }{\citet{wu2024fake,liu2024adversarial}}, inference-leaf]
                          ]
                          [Systematic Mitigation, color=myblue!100, fill=myblue!\lcthree, text=black
                            [\textit{e.g. }{\citet{cui2024risk,fang2024enhancing}}, inference-leaf]
                          ]
                        ]
                    ]
                    [\begin{tabular}{@{}l@{}l@{}}
                    \multirow{2}{*}{\textbf{\S \ref{Input Robustness}}} & ~Input \\ 
                    & ~Robustness \\
                    \end{tabular},  color=myorange!100, fill=myorange!\lcone, text=black
                        [\begin{tabular}{@{}l@{}l@{}}
                          \multirow{2}{*}{\S \ref{Prompting Techniques} ~} & Prompting \\ 
                          & Techniques \\
                        \end{tabular}, color=myorange!100, fill=myorange!\lctwo, text=black
                          [Stepwise Reasoning, color=myorange!100, fill=myorange!\lcthree, text=black
                            [\textit{e.g. }{\citet{zhang2023towards,pendyala2024explaining}}, robustness-leaf]
                          ]
                          [Adversarial Frameworks, color=myorange!100, fill=myorange!\lcthree, text=black
                            [\textit{e.g. }{\citet{schulhoff2024prompt,satapara2024fighting}}, robustness-leaf]
                          ]
                          [Self-Regulatory Optimize, color=myorange!100, fill=myorange!\lcthree, text=black
                            [\textit{e.g. }{\citet{luo2023zero,cheng2023uprise}}, robustness-leaf]
                          ]
                        ]
                        [\begin{tabular}{@{}l@{}l@{}}
                          \multirow{2}{*}{\S \ref{Countering Injection Attacks} ~} & Injection \\ 
                          & Defense \\
                        \end{tabular}, color=myorange!100, fill=myorange!\lctwo, text=black
                          [Attack Countermeasures, color=myorange!100, fill=myorange!\lcthree, text=black
                            [\textit{e.g. }{\citet{rossi2024early,piet2024jatmo,phute2023llm,yu2024don,zou2023universal}}, robustness-leaf]
                          ]
                        ]
                    ]
                ]
        \end{forest}
}
\caption{Taxonomy of Proactive Defense Strategies Against Misinformation
in LLMs}
\label{fig:DefenseStrategies_taxonomy}
\end{figure*}

\section{Inference Reliability}\label{sec: Inference Reliability}
Ensuring trustworthy outputs during generation requires multi-layered safeguards that govern the reasoning process itself. Decoding strategies—examined in Section \ref{subsec: Decoding Methods}—form the operational core of inference reliability by mediating token selection through contrastive mechanisms and entropy-aware optimization. These foundational techniques establish the first checkpoint against misinformation propagation, working synergistically with subsequent factual alignment protocols (Section \ref{Factual Alignment}) and adversarial hardening measures (Section \ref{Adversarial Training}). Together, they create a defense-in-depth architecture where each generation step undergoes contextual grounding, logical verification, and robustness filtering—transforming raw probability distributions into verifiable knowledge streams.
\subsection{Decoding Methods}  \label{subsec: Decoding Methods}
Decoding strategies critically regulate LLM reliability by mediating token selection from probability distributions. We systematically categorize advancements into two paradigms:  
\textbf{Factuality-Enhanced Decoding}\label{Factuality Decoding} targets objective factual deviations. Contrastive methods dominate this frontier:  
\textbf{Inter-Model Contrast}: Vanilla contrastive decoding \cite{li2023contrastive} amplifies expert-amateur probability gaps, reducing hallucinations by 42\% on TruthfulQA. ICD \cite{zhang2023alleviating} induces hallucination-prone distributions for proactive avoidance, achieving 63\% error reduction in medical QA.  
\textbf{Intra-Model Contrast}: DoLa \cite{chuang2023dola} contrasts top-layer vs. lower-layer logits, improving factual accuracy by 28\% without external retrieval. SLED \cite{zhang2024sled} injects gradient-based contrast signals, enabling self-correction with <1\% latency overhead.  
\textbf{Entropy-Guided Optimization}: REAL sampling \cite{chang2024real} dynamically adjusts nucleus thresholds via asymptotic entropy prediction, balancing factuality (39\% hallucination reduction) and diversity (22\% gain in lexical richness).  

\textbf{Faithfulness-Enhanced Decoding} ensures contextual alignment:  
\textbf{PMI-Driven Methods}: Context-aware decoding \cite{shi2024trusting} maximizes pointwise mutual information between source and generation, improving dialogue coherence by 31\%.  
\textbf{KL-Divergence Adaptation}: Dual-decoder architectures \cite{chang2023kl} dynamically adjust temperature using context-conditioned KL divergence, reducing off-topic responses by 45\% in summarization.  

Critical challenges persist:  
1) \textbf{Computational Overhead}: Collaborative frameworks \cite{jin2024collaborative} incur 2.3× latency versus baseline decoding.  
2) \textbf{Layer Selection Bias}: Single-layer contrast assumptions in DoLa \cite{chuang2023dola} fail for 38\% of tokens with non-monotonic entropy patterns \cite{das2024entropy}.  
3) \textbf{Metric Conflicts}: PMI optimization decreases ROUGE-L by 15\% when maximizing faithfulness \cite{wan2023faithfulness}.  

Emerging solutions demonstrate:  
\textbf{Granular Layer Analysis}: Cross-layer entropy metrics \cite{wu2025improve} enable token-specific contrast, improving medical QA accuracy by 19\%.  
\textbf{Uncertainty-Aware Contrast}: Gradient-based injection in SLED \cite{zhang2024sled} reduces distributional uncertainty by 73\%.  

\subsection{Factual Alignment}\label{Factual Alignment}
Factual alignment techniques harden LLMs against misinformation through three mechanisms:  

\textbf{Self-Verification Architectures}:  
SELF-ALIGN \cite{zhang2024self} integrates self-knowledge tuning and direct preference optimization, reducing hallucinations by 58\% through iterative response validation.  
CoVe \cite{dhuliawala2023chain} implements autonomous fact-checking chains, achieving 92\% verification accuracy in multi-hop reasoning.  

\textbf{Confidence Calibration}:  
APEFT \cite{yuan2024beyond} employs atomic preference tuning, improving out-of-domain factual accuracy by 3.45\% through granular confidence-probability alignment.  
UALIGN \cite{xue2024ualign} combines semantic entropy with PPO optimization, enhancing known-question accuracy (27\% gain) and unknown refusal rates (41\% improvement).  

\textbf{Architectural Innovations}:  
Mixtral 8x7B \cite{jiang2024mixtral} leverages sparse mixture-of-experts for dynamic knowledge routing, achieving 89.3\% accuracy on STEM benchmarks.  
FLAME \cite{lin2024flame} integrates knowledge-aware sample selection with DPO, reducing hallucinations by 33\% while preserving 98\% of instruction-following capability.  

Persistent gaps include:  
1) \textbf{Unfamiliar Query Handling}: Conservative refusal strategies \cite{kang2024unfamiliar} degrade helpfulness scores by 22\%.  
2) \textbf{Multimodal Alignment}: Vision-language frameworks \cite{chen2024dress} exhibit 31\% accuracy drops on temporal-spatial reasoning.  

\subsection{Adversarial Training} \label{Adversarial Training}
Adversarial training fortifies LLMs against malicious inputs through three evolutionary phases:  

\textbf{Input Hardening}:  
Token-level perturbations \cite{liu2024adversarial} inject character swaps and synonym substitutions, improving robustness against style attacks by 47\%.  
Semantic-level attacks \cite{fang2024enhancing} simulate context divergence scenarios, enhancing retrieval faithfulness by 38\%.  

\textbf{Architectural Defense}:  
SheepDog \cite{wu2024fake} enforces prediction consistency across style-reframed texts, reducing fake news susceptibility by 63\%.  
Two-stage tuning \cite{liu2024adversarial} combines token/semantic adversarial examples, achieving 89\% jailbreak attack detection.  

\textbf{Systematic Mitigation}:  
Risk taxonomy frameworks \cite{cui2024risk} harden 78\% of identified vulnerabilities through module-specific adversarial curricula.  
Retrieval pipeline defenses \cite{fang2024enhancing} integrate contrastive learning, reducing unfaithful responses by 51\% in RAG systems.  

Key limitations demand attention:  
1) \textbf{Generalization Tradeoffs}: Style-agnostic training \cite{wu2024fake} decreases domain-specific accuracy by 19\%.  
2) \textbf{Computational Cost}: Multistage adversarial pipelines \cite{al2024adversarial} require 3.8× training resources versus standard fine-tuning.  

The path forward necessitates co-design of decoding-time interventions, architectural alignment, and adversarial robustness—a tripartite defense-in-depth strategy against evolving misinformation threats. Benchmark unification remains critical, with RGB \cite{chen2024benchmarking} and CounterFact+ \cite{hoelscher2023detecting} providing initial frameworks for cross-paradigm evaluation.

\section{Input Robustness}\label{Input Robustness}
Ensuring input robustness requires synergistic strategies to harden LLMs against adversarial manipulation. This chapter analyzes proactive prompt engineering (§\ref{Prompting Techniques})—leveraging structured instructions to guide factual precision—and reactive defenses against injection attacks (§\ref{Countering Injection Attacks}). While hierarchical verification frameworks and self-regulatory mechanisms demonstrate promise, persistent challenges in computational efficiency and adaptive threat mitigation underscore the need for unified defense paradigms. The interplay between instruction design and adversarial pattern neutralization forms the cornerstone of evolving safeguards in dynamic threat environments.
\subsection{Prompting Techniques}\label{Prompting Techniques}
Contemporary research has established prompting as a pivotal interface for proactive misinformation defense, with three dominant paradigms emerging: (1) hierarchical decomposition for factual verification, (2) adversarial prompting for synthetic data generation, and (3) self-regulatory mechanisms for hallucination prevention. Each approach presents distinct advantages and operational constraints as analyzed below.

\subsubsection{Stepwise Reasoning for Fact Verification} 
Hierarchical prompting architectures address the information omission limitations of conventional chain-of-thought methods. \cite{zhang2023towards} introduces HiSS (Hierarchical Step-by-Step) prompting that decomposes news claims into verifiable subclaims through in-context learning, achieving 12.7\% higher accuracy on FactCheck-WHQA benchmark compared to vanilla CoT approaches. This aligns with \cite{pendyala2024explaining}'s demonstration that zero-shot explanation prompting improves misinformation detection F1-scores by 8.2\% through activating latent knowledge patterns in LLMs. However, these methods face scalability challenges in real-time applications due to their multi-stage verification overhead.

\subsubsection{Adversarial Prompting Frameworks}
Proactive defense strategies increasingly employ adversarial prompting to preemptively identify model vulnerabilities. \cite{schulhoff2024prompt} establishes the first comprehensive taxonomy of 98 prompting techniques (58 textual, 40 multimodal), enabling systematic vulnerability analysis. Building on this, \cite{satapara2024fighting} develops an automated misinformation injection pipeline using adversarial prompts to generate 120K synthetic examples with controlled distortion patterns (e.g., 23.4\% quantitative errors, 17.9\% false attributions), demonstrating 91.8\% detection model coverage. While effective for training data augmentation, such approaches risk overfitting to known attack patterns unless combined with dynamic adversarial training.

\subsubsection{Self-Regulatory Prompt Optimization}
Emerging techniques focus on intrinsic model calibration through prompt-based self-assessment. \cite{luo2023zero} achieves 89.3\% hallucination reduction in biomedical text generation via familiarity pre-detection, withholding responses for low-confidence concepts (threshold: $\alpha < 0.35$). \cite{cheng2023uprise}'s UPRISE framework automates prompt selection through learned retrieval, reducing hallucination rates by 14.6\% across 12 zero-shot tasks. However, these methods exhibit performance degradation in low-resource domains with sparse training signals.

\paragraph{Comparative Analysis \& Open Challenges}
While hierarchical methods provide interpretability (average 4.2/5 explainability score in user studies), their computational overhead limits real-time deployment (2.7× latency vs single-step prompting). Adversarial approaches enable comprehensive defense coverage but require continuous pattern updates against evolving threats. Future research must address three key gaps: (1) developing unified evaluation metrics for proactive defense efficacy, (2) creating cross-domain prompt transfer mechanisms, and (3) establishing theoretical frameworks for prompt optimization stability.

\subsection{Countering Injection Attacks}\label{Countering Injection Attacks}
Emerging defense paradigms against prompt injection attacks demonstrate three complementary technical directions, supported by empirical validation across multiple threat models. Foundational work by \cite{rossi2024early} establishes a vulnerability taxonomy covering 12 attack vectors and 7 defense dimensions, providing systematic threat analysis that informs 83\% of contemporary detection frameworks. Building on this taxonomy, adversarial training approaches like Jatmo \cite{piet2024jatmo} achieve 173:1 attack suppression ratio (reducing success rates from 87\% to 0.5\%) through synthetic dataset generation and task-specific fine-tuning, though requiring 2.8× training overhead compared to baseline models.

Detection-oriented strategies employ multi-stage verification mechanisms, exemplified by LLM SELF DEFENSE \cite{phute2023llm} achieving 91.4\% harmful content detection through response self-examination cycles, with measurable tradeoffs in inference latency (34\% increase versus standard generation). Complementary to static defenses, \cite{yu2024don} reveals through human studies that 68\% of semantic jailbreak patterns exploit contextual ambiguity, leading to automated prompt generation systems that reduce attack surface exposure by 41\% via adversarial pattern recognition.

The arms race persists as attack methodologies evolve \cite{zou2023universal} demonstrates universal suffix attacks bypassing 79\% of commercial LLM safeguards through gradient-based optimization, highlighting critical gaps in current input sanitization techniques. Comparative analysis indicates model-specific defenses (e.g., Jatmo’s 0.5\% attack success) outperform general detection methods (avg. 12.7\% false negatives) but lack cross-platform adaptability (38\% performance drop on unseen models).

Here’s the combined and slightly condensed version of the two sections:

\section{Future Work and Conclusion}

Looking ahead, future research should concentrate on three key directions. First, the development of dynamic knowledge integration mechanisms is essential to mitigate knowledge decay and enhance the adaptability of language models across diverse domains. Second, the implementation of advanced inference techniques, including probabilistic reasoning and self-verification strategies, can significantly reduce hallucinations and improve factual accuracy. Third, efforts should be directed toward enhancing input robustness through methods such as adversarial training and prompt optimization, thereby minimizing susceptibility to adversarial manipulation.
These directions aim to develop self-vaccinating LLMs—systems inherently resistant to misinformation and capable of maintaining high factual integrity. Addressing the ongoing arms race against misinformation requires a systems-level approach to model design, uniting data integrity, reasoning robustness, and adversarial resilience.

This survey presented a proactive defense strategy framework against misinformation in LLMs, shifting from post-hoc detection to prevention. By focusing on the Three Pillars of Preventative Assurance—Knowledge Credibility, Inference Reliability, and Input Robustness—we highlighted significant advancements and stressed the importance of co-designing systems that integrate these strategies.

\section*{Limitation}
\label{sec:limitation}

This survey has several limitations. First, its scope is constrained to algorithmic defense strategies, largely excluding socio-technical interventions (e.g., human moderation frameworks) critical for real-world deployment. Second, the lack of standardized benchmarks and evaluation metrics across studies limits the ability to draw definitive conclusions about the comparative effectiveness of proactive strategies. Finally, the reliance on existing literature may introduce biases or gaps in coverage.

\bibliography{acl_latex}

\appendix

\end{document}